# Lattice Dynamics, Mechanical Properties, Electronic Structure and Magnetic Properties of Equiatomic Quaternary Heusler Alloys CrTiCoZ (Z =Al,Si) using *first principles* calculations

**Eesha Andharia** [a,b], **Hind Alquarashi** [b,c] and **Bothina Hamad** [a,b,d,*]

[a]Department of Physics, University of Arkansas, Fayetteville, AR 72701, USA
[b]Materials Science and Engineering Program, University of Arkansas, Fayetteville, AR 72701, USA
[c]Physics Department at the College of Science, Al Baha University, Al Baha 65527, Saudi Arabia
[d]Physics Department, The University of Jordan, Amman-11942, Jordan
* Correspondence: Corresponding Author: bothinah@uark.edu

**ABSTRACT:** First principles calculations are performed to investigate the thermodynamical stability, dynamical, mechanical, electronic and magnetic properties of CrTiCoZ (Z= Al/Si) novel quaternary Heusler alloys. Y-type III atomic configuration is found to be the most stable structure for both compounds. The lattice constant values obtained using GGA-PBE approach are 5.9368Å and 5.7853Å for CrTiCoAl and CrTiCoSi, respectively. Using the value of elastic moduli for both the compounds, the computed Pugh's ratio value is 2.5 and 2.7 for CrTiCoAl and CrTiCoSi, respectively, which is higher than 1.75, indicating both the compounds are ductile in nature. The melting temperatures of both compounds are as high as 2142K and 2420K for CrTiCoAl and CrTiCoSi, respectively. The electronic structure calculations using GGA-PBE approach show a half metallic behavior of CrTiCoAl. The spin-down channel exhibits a direct band gap of 0.15 eV, whereas the spin-up channel is metallic making CrTiCoAl a half metallic ferromagnet with 100% spin polarization and an appreciable magnetic moment of -2$\mu_B$. However, CrTiCoSi is found to be semi-metallic in the spin-down channel and metallic in the spin-up channel, which leads to a spin polarization of 99.7% with a non-integer magnetic moment of -0.99$\mu_B$. The Curie temperature of CrTiCoAl is well above the room temperature (385K), whereas that of CrTiCoSi is below the room temperature (203K). Thus, CrTiCoAl is found to be more promising than CrTiCoSi as a spin injector in spintronic devices.

**Keywords:** Equiatomic Quaternary Heusler alloys (EQHA); phonons; Half-metallic ferromagnet; Density Functional Theory; spin polarization; spintronics

## 1. Introduction

Spintronics is a rapidly progressing branch of nanoelectronics that relies on manipulation of the spin of an electron rather than its charge for data storage and transfer as well as information processing [1]. The basic building block of spintronics devices is called a magnetic tunnel junction (MTJ), which consists of two ferromagnetic electrodes separated via an insulating spacer or barrier. One of the ferromagnetic (FM) electrode has a fixed magnetization direction and is known as pinned or reference layer whereas the magnetization of the other FM electrode can be switched via external current or magnetic fields and is known as free layer. These devices operate on the principle of quantum tunneling and spin transfer torque (STT) [2]. The change from parallel to anti-parallel alignment of the free layer with reference to the pinned layer gives rise to a very high tunnel magneto-resistance (TMR) in these devices. The TMR value is related to the spin polarization of reference and free ferromagnetic electrodes by the equation [3].

$$TMR = \frac{2P_1 P_2}{1 - P_1 P_2} \qquad (1)$$

where, $P_1$ and $P_2$ are spin polarizations of the ferromagnetic electrode 1 and 2, respectively. To obtain a high TMR ratio at room temperatures, the ferromagnetic electrodes must have high spin-polarization (ideally 100%). Half-metallic ferromagnets (HMF) are metallic in one spin channel and semiconducting or insulating in the other spin channel. Hence, they have a 100% spin polarization, which makes them ideal candidates for spintronic applications. Examples of theoretically predicted conventional half-metallic ferromagnetic materials include chalcogenides like $Ag_3CrX_4$ (X = S, Se, and Te) [4] and $Cu_3TmCh_4$ (Tm = Cr,Fe and Ch = S,Se,Te)[4, 5], $CrO_2$[6] and $Fe_3O_4$[7]metallic oxides, $PrMnO_3$[8] and $Ce(Fe/Cr)O_3$[9] perovskites, and Heusler alloys[10]. Amongst these compounds, Heusler alloys represents a family of a wide variety of compounds that can be easily fabricated and possess high Curie temperature making them ideal candidates for MTJs fabrication [11].

Heusler alloys can be widely classified into ternary or quaternary alloys. Ternary Heusler alloys can be further classified into half Heusler compounds with chemical formula XYZ and full Heusler alloys $X_2YZ$, which crystallize in C1$_b$ and L2$_1$ structures, respectively. Here, X and Y represent transition metal elements or Lanthanides and Z is main group element. On the other hand, quaternary Heusler alloys with a chemical formula XX'YZ and 1:1:1:1 stoichiometry consist of four interpenetrating FCC lattices with space group number 216 (F$\bar{4}$3m). Half metallic ferromagnetism was first discovered in ternary half-Heusler alloy NiMnSb [12, 13] which had a GMR value of only 1% at 60K[14]. Later in 2000, it was observed that a transition from half-metallic ferromagnet to metallic ferromagnetic behavior in NiMnSb occurs at 80K, well below the Curie temperature of NiMnSb[15]. CoTiSb and NiTiSn were also predicted to be half-metallic ferromagnets, which exhibited a similar transition to metallic state by changing the concentration of valence electrons[16, 17]. Recently, half-metallic ferromagnetism was predicted in ternary half-Heusler compounds like NiCrZ(Z =Si, Ge, Ga, Al, In, As)[18], XCrSb (X = Ti, Zr, Hf) [19], RuMnZ (Z = P, As) [20], GeNaZ (Z = Sr, Ba andCa)[21] and CoTcSn[22]. The half metallic ferromagnetism was predicted in several examples of ternary full-Heusler alloys including $Co_2FeSi$ [23], $Co_2MSn$ (M = Ti, Zr, Hf)[24], $Ti_2FeSn$[25], $Co_2YZ$ (Z = P, As, Sb, Bi)[26] and $Co_2YGe$ (Y = Mn, Fe) [27]. However, equiatomic quaternary Heusler alloys have an advantage of low power dissipation due to their negligible disordered scattering as compared to ternary Heuslers [28]. Some of these compounds that are spin gapless semiconductors (SGS) (semi-metallic in one spin channel and semiconducting in other spin channel) also lead to a 100% spin polarization. Several examples of theoretically predicted half metallic ferromagnetic EQHAs include CoFeCrZ (Z = Al,Ga,Si)[29], CoYCrZ ( Z = Si,Ge,Ga,Al)[30], CoFeMnZ (Z = Al,Ga,Si, Ge)[31], CoFeCrZ (Z = P, As, Sb)[32], CoRhMnZ (Z=Al, Ga,Ge,Si)[33], CoMnTiZ (Z=P,As,Sb)[34], CoCuMnZ (Z = In,Sn,Sb)[35], CoRhMnGe[36], CoCrIrSi[37]. There are only few experimental studies using EQHAs in MTJs. Among these are CoFeMnSi [38] and CoFeCrAl[39], which have high TMR ratios of 101% and 87% at room temperature, respectively. Further, there is a lot of room left to theoretically investigate several new combinations of HMF or SGS EQHA.

Özdoğan et al [40] investigated the electronic and magnetic properties of 60 different EQHAs, where CoCrTiAl and CoCrTiSi are two combinations amongst them. However, this study is limited as they have not identified the most stable atomic configurations and have not checked for the dynamical and mechanical stability of these compounds. In this paper, we adopt a holistic approach to check the pragmatic viability of CrTiCoAl and CrTiCoSi EQHAs for spintronics applications using density functional theory. In the first part, we check for its thermodynamical stability by calculating the formation energies of the most stable configuration followed by checking the mechanical stability of these compounds by calculating their stress-strain relations. We further bolstered our results by studying the dynamical phonons for both compounds. In the second part, we discuss the electronic and magnetic properties of the most stable Y-type III atomic configurations of these compounds. Apart from electronic band-structure and density of states, we studied their spin polarization and net magnetic moments using Full Potential – Linearized Augmented Plane wave (FP-LAPW) method.

## 2. Computational methods

The formation energies, mechanical stability and phonon calculations were performed using spin-polarized density functional theory as implemented in Vienna ab-initio simulation package (VASP) [41, 42] using projected augmented wave (PAW) type of pseudopotentials [43]. The exchange-correlation term was treated using generalized gradient approximation (GGA) as implemented by Perdew-Burke-Ernzerhof (PBE) [44]. The plane-wave energy cut-off was set to 400 eV and the energy-convergence criterion was set to $10^{-8}$ eV. A 10×10×10 K-mesh was used for the Brillouin Zone (BZ) integration. Then the dynamical stability of the two stable compounds was verified using Phonopy package [45] using finite displacement method (FDM) [46, 47] with a 4×4×4 supercell and a 4×4×4 Γ-centered k-mesh. As above-mentioned, the electronic and magnetic properties were studied using WIEN2k code [48] with an $R_{MT}K_{max}$ value of 7, where $R_{MT}$ is the minimum value of muffin-tin sphere radius and $K_{max}$ is the largest value of reciprocal lattice vector in the first BZ. The charge convergence was set to $10^{-4}$ Ry. The $R_{MT}$ values were set to 2.3, 2.3, 2.3, 2.16 and 1.91 atomic units (a.u.) for Cr, Ti, Co, Al and Si, respectively. The core states were separated from the valence states with an energy cutoff of -6 Ry.

## 3. Results and discussion

The thermodynamic, structural, dynamical and mechanical properties are presented in section 3.1, whereas the electronic and magnetic properties of the most stable structure types are in section 3.2.

### 3.1. Structural, dynamical and mechanical properties

These compounds crystallize in Y-type LiMgPdSn structure, which has three different atomic configurations: Y-type-I, II and III as shown in Figure 1. (Wyckoff positions of all the atoms are stated in Table 1).

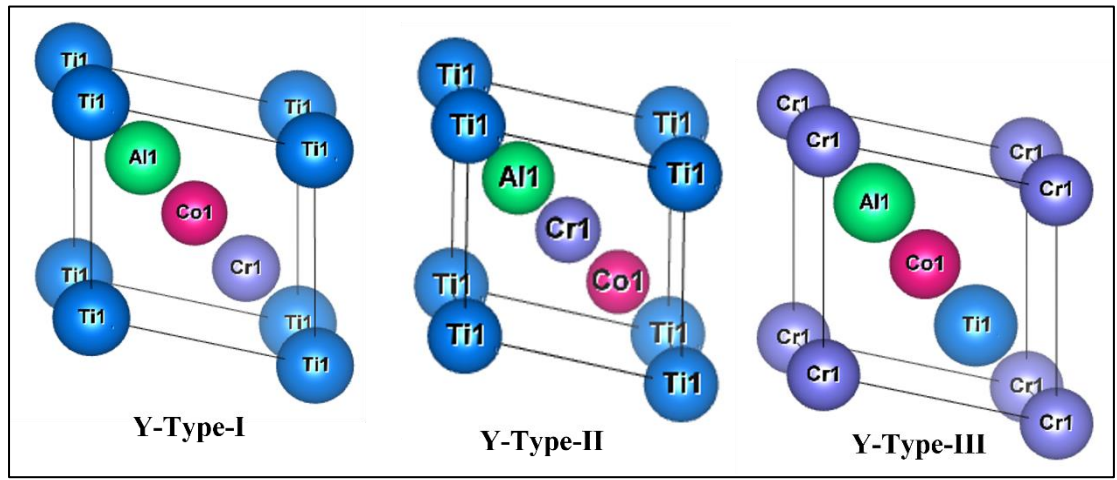

**Figure 1.** Primitive unit cells of CrTiCoAl quaternary Heusler alloys in the Y-Type-I,. Y-Type-II and Y-Type-III configurations.

**Table 1.** Wyckoff positions of the atoms in different configurations (Z stands for Al or Si).

| Y | 4a (0,0,0) | 4c ($\frac{1}{4},\frac{1}{4},\frac{1}{4}$) | 4b ($\frac{1}{2},\frac{1}{2},\frac{1}{2}$) | 4d ($\frac{3}{4},\frac{3}{4},\frac{3}{4}$) |
|---|---|---|---|---|
| Type-I | Ti | Cr | Co | Z |
| Type-II | Ti | Co | Cr | Z |
| Type-III | Cr | Ti | Co | Z |

Structural optimization was performed in VASP using the conjugate gradient approximation to obtain the lattice constants in all three configurations for both compounds. For CrTiCoAl, the optimized lattice parameters for Y-type-I, II and III structures are 6.0068 Å, 6.0623 Å and 5.9368 Å, respectively. Similarly for CrTiCoSi, the optimized lattice parameters are 5.9126 Å, 5.9505 Å and 5.7853 Å for Y-type-I, II and III, respectively. The next step in determining the structural stability was to calculate the formation energies for all three Y-types structures as per equation 2 to find the thermodynamically most stable atomic configuration.

$$E_{formation} = E_{total} - (E_{Ti}^{bulk} + E_{Cr}^{bulk} + E_{Co}^{bulk} + E_{Z}^{bulk}) \qquad (2)$$

Here, $E_{total}$ stands for the total ground state energy of the primitive cell, whereas, $E_{Ti}^{bulk}$, $E_{Cr}^{bulk}$, $E_{Co}^{bulk}$ and $E_{Z}^{bulk}$ are the ground state energies of Ti, Cr, Co and Z=Al or Si, per formula unit, respectively. While computing the ground state energies per formula unit of individual bulk Ti, Cr, Co, and Si, their unit cells were considered to crystallize in hcp, bcc, hcp, bcc and diamond cubic structures, respectively. The individual ground state energies for Ti, Cr, Co,Al and Si are -7.7755eV, -9.5075eV, -7.10928eV, -3.7363eV and -5.4234eV, respectively. Table 2 shows the ground state energies, E$_{total}$ for both the compounds in all three atomic configurations.

**Table 2.** Ground state energies of both compounds in the three atomic configurations, in eV.

| Alloys | Y-Type-I | Y-Type-II | Y-Type-III |
|---|---|---|---|
| CrTiCoAl | -28.3434 | -28.5583 | -29.2646 |
| CrTiCoSi | -30.3996 | -30.7391 | -31.4684 |

As shown in Table 3, the most favorable structure with the minimum formation energies corresponds to Y-type structure-III for both the cases of CrTiCoAl and CrTiCoSi compounds. From here onwards all calculations will adopt these two stable compounds.

**Table 3.** Formation energies of both compounds in the three atomic configurations, in eV.

| Alloys | Y-Type-I | Y-Type-II | Y-Type-III |
|---|---|---|---|
| CrTiCoAl | -0.21 | -0.42 | -1.13 |
| CrTiCoSi | -0.58 | -0.92 | -1.65 |

The dynamical stability is further checked for these compounds by calculating the phonon dispersion curves for both alloys. Within the harmonic approximation, the dynamical properties are obtained by solving the following eigen-value equation:

$$D(q)e_{qj} = \omega_{qj}^2 e_{qj} \qquad (3)$$

Where, D(q) is the dynamical matrix calculated as below:

$$D_{kk'}^{\alpha\beta}(q) = \sum_{l'} \frac{\Phi_{\alpha\beta}(ok,l'k')}{\sqrt{m_k m_{k'}'}} e^{iq.[r(l'k'-r(ok)} \qquad (4)$$

Where, $m_k$ is the mass of the atom, $\omega_{qj}$ is phonon frequency and $e_{qj}$ is polarization vector and $\Phi_{\alpha\beta}$ is second order force constant (l and k are labels of unit cells and atoms, respectively). A crystal is dynamically stable if the potential energy increases for any atomic displacements. Within the harmonic approximation this is translated to positive and real frequencies in phonon dispersion relations.

As shown in Figure 2, both compounds have only positive frequencies in the most stable atomic configurations, implying that both are dynamically stable. These results are in good agreement with references [49, 50]. A single unit cell consists of 4 atoms, hence, there are 3N (N=4), which correspond to 12 phonon branches out of which first three at lower frequencies represent the acoustic phonon modes, while the rest at higher frequencies are optical modes. One of the three is longitudinal acoustic (LA), while the other two are transverse acoustic (TA) modes. In addition, it is found that the two transverse acoustic branches are degenerate from L to Γ and Γ to X points and hence they appear as a single branch along this path for compounds. It is seen from the projected-phonon density of states that for CrTiCoAl, the low frequency acoustic branches are formed by a major contribution from Ti atoms whereas at higher optical frequencies Cr and Al atoms are dominating. Similarly, for CrTiCoSi, the major contribution to the lower frequency branches are coming from Co atoms whereas the higher frequency branches are formed of contributions from Cr and Si atoms.

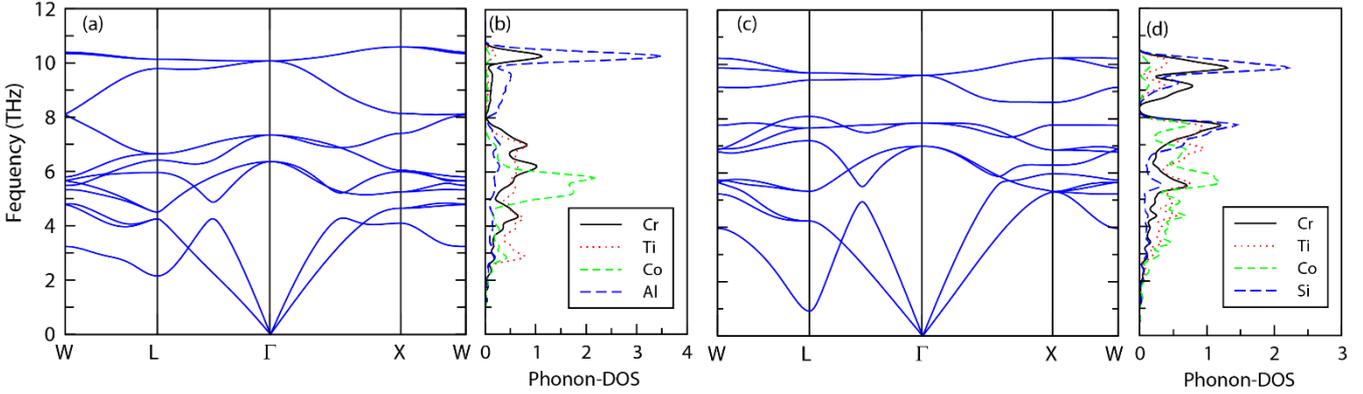

**Figure 2.** Phonon dispersion relations of (a) CrTiCoAl and (c) CrTiCoSi and Projected phonon density of states for (b) CrTiCoAl and (d) CrTiCoSi

These results are further bolstered by investigating the mechanical properties of both compounds. Mechanical properties are obtained by calculating the fourth order elastic moduli tensor. The stress-strain relationship is derived from the calculation of Hessian matrix (second derivative of energy with respect to atomic positions) by performing a total of 6 atomic distortions of the lattice. Since these compounds crystallize in a cubic structure, each has three different elastic constants: longitudinal compression - $C_{11}$, transverse expansion – $C_{12}$ and shear modulus component – $C_{44}$ as listed in Table 4. This table shows that the elastic constants of both compounds satisfy the Born-Huang conditions [51] as stated in equation 5 and hence are mechanically stable.

$$\frac{C_{11} - C_{12}}{2} > 0; \frac{C_{11} + 2C_{12}}{3} > 0; \; C_{44} > 0 \tag{5}$$

Various other mechanical properties like bulk modulus (B), Voigt ($G_V$), Reuss ($G_R$) shear modulus (G), Young's modulus (E), Cauchy pressure ($C_P$), Pugh's ratio (B/G), and anisotropy factor (A) can be calculated using the relations (6) –(12) [52-54] and are listed in Table 4.

$$B = \frac{(C_{11} + 2C_{12})}{3} \tag{6}$$

$$G = \frac{(G_v + G_r)}{2} \tag{7}$$

$$G_v = \frac{C_{11} - C_{12} + 3C_{44}}{5} \tag{8}$$

$$G_R = \frac{(5C_{44}(C_{11} - C_{12}))}{4C_{44} + 3(C_{11} - C_{12})} \tag{9}$$

$$E = \frac{9GB}{3B + G} \tag{10}$$

$$C_p = \; C_{12} - C_{44} \tag{11}$$

$$A = \frac{2C_{44}}{C_{11} - C_{12}} \qquad (12)$$

Bulk Modulus (B) and Shear modulus (G) are used to calculate the Pugh's ratio (B/G), which is an indication of the ductility of materials. The obtained values of Pugh's ratio are 2.5 and 2.7 for CrTiCoAl and CrTiCoSi, respectively. These values are higher than 1.75, which indicates that both alloys are ductile in nature. This result is further supported by the positive values of Cauchy pressure $C_P$ indicating the ductility of both compounds. The value of Young's modulus, E, is 173.1 GPa for CrTiCoAl and 204.4 GPa for CrTiCoSi, which indicates that CrTiCoSi is stiffer than CrTiCoAl. A value of 1 for the anisotropic factor, A, means that the compounds are isotropic. However, in our case the value of A is smaller than 1 indicating that both compounds are anisotropic. These results are similar to the method followed in references [55-57] for calculating the mechanical properties. The values of the formation energies as well as mechanical properties calculated above are similar to those of other theoretically predicted EQHAs like CoRhMnZ(Z = Al, Ga, Ge and Si) [33], CoFeCrZ (Z = P, As and Sb)[32], CoYCrZ (Z = Si, Ge, Ga, Al) [30] and CoCrIrSi [37]. The melting temperature for both compounds is calculated using the following equation [58, 59]:

$$T_{melt} = \left[ 553K + \left(\frac{5.91K}{GPa}\right)C_{11} \right] \pm 300K. \qquad (13)$$

The above equation is based on empirical relation studies by Fine *et. al.* where, $T_{melt}$ increases with increasing $C_{11}$. This behavior is similar to those predicted for other metallic cubic crystals[60]. Other examples of intermetallic compounds obeying such a relationship are provided in reference [61] . As evident from Table 4, since the value of the longitudinal compression, $C_{11}$, is lower for CrTiCoAl than that of CrTiCoSi, its melting temperature (2142K) is also lower than that of CrTiCoSi (2420 K). The melting temperatures calculated here are analogous to those of other EQHA compounds such as VTiRhSi (2296K) [62], VTiRhGe (2053K) [62], CoFeCrGe (2584K) [62], and CoFeTiGe (2484K) [63].

**Table 4.** Lattice constant and mechanical properties of Y-Type-III stable atomic configurations of CrTiCoAl and CrTiCoSi compounds.

|  | CrTiCoAl | CrTiCoSi |
|---|---|---|
| a (Å) | 5.9368 | 5.7853 |
| $C_{11}$ (GPa) | 268.8 | 316.0 |
| $C_{12}$(GPa) | 115.4 | 152.2 |
| $C_{44}$(GPa) | 58.6 | 73.1 |
| B(GPa) | 166.6 | 206.8 |
| G(GPa) | 65.3 | 76.6 |
| E(GPa) | 173.2 | 204.4 |
| B/G | 2.55 | 2.70 |
| $C_P$(GPa) | 56.8 | 79.0 |
| A | 0.76 | 0.89 |
| $T_{melt}$ (K) | 2142 | 2420 |

### 3.2. Electronic and Magnetic properties

In this section we present the electronic properties, namely the band structure, total density of states (TDOS), and projected density of states (PDOS). In addition, the magnetic properties namely, the net magnetic moment, spin polarization and Curie temperature for both compounds are studied.

Figures 3 and 4 show the electronic band structure as well as total density of states (TDOS) for both alloys. As shown in Fig. 3, CrTiCoAl alloy exhibits a metallic behavior in the spin-up channel, whereas it is semiconducting with a direct band gap of 0.15 eV in the spin-down channel. Unlike CrTiCoAl, the spin-down channel of CrTiCoSi exhibits a semi-metallic behavior, whereas the spin-up channel is metallic (Figure 4). The results of CrTi-CoAl alloy are in good agreement with a previous investigations of VTiRhZ (Z = Si,Ge,Sn)[62] that show a half metallic ferromagnetic behavior.

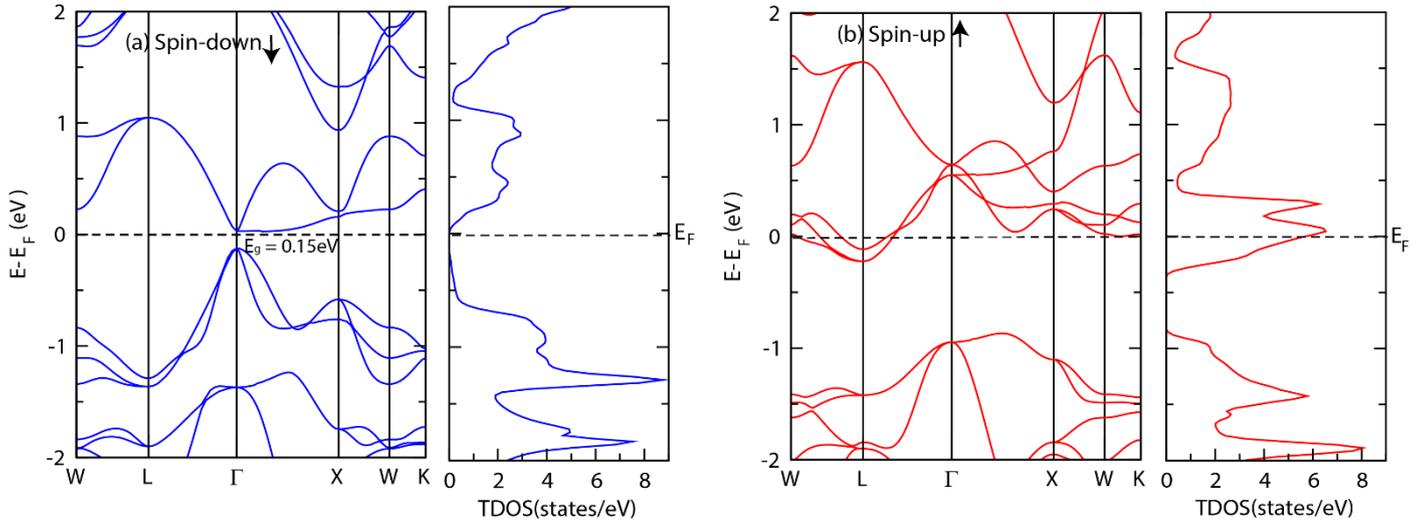

**Figure 3.** Electronic band structure and total density of states (TDOS) of CrTiCoAl in (a) the spin up channel and (b) the spin down channel.

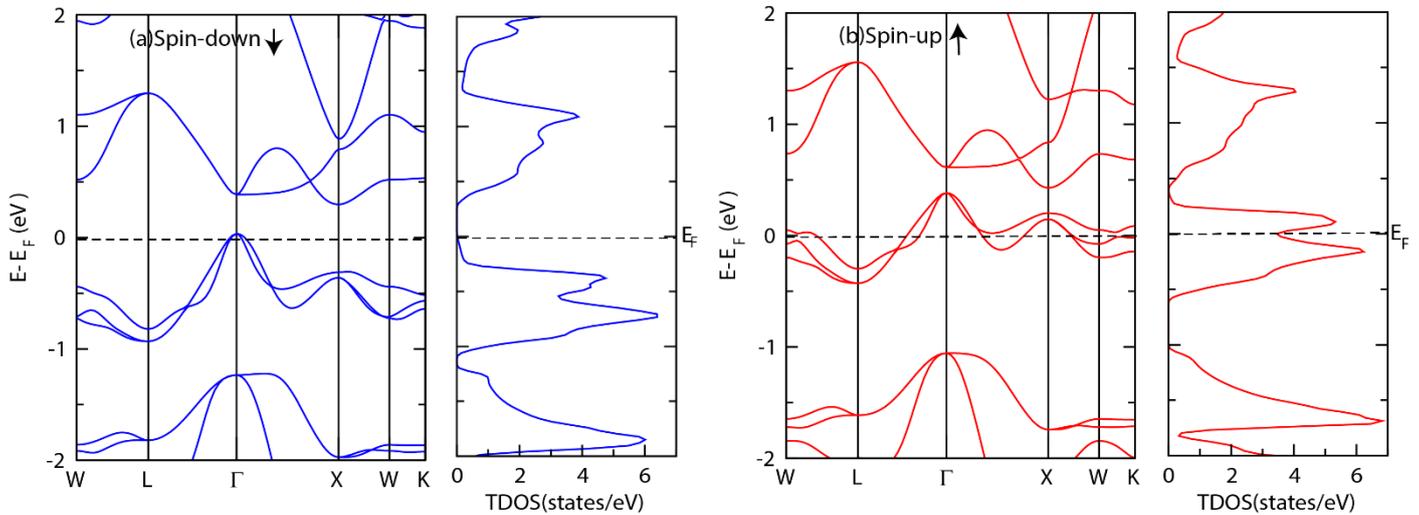

**Figure 4.** Electronic band structure and total density of states (TDOS) of CrTiCoSi in (a) The spin up channel and (b)the spin down channel.

Figures 5a and 5b present the projected density of states (PDOS) of the spin-up and spin-down channels for CrTiCoAl and CrTiCoSi alloys. The main contribution to the top of the valence band comes from Cr-*d* orbital, although, *d* orbitals of Co and Ti atoms also have appreciable contributions. However, the bottom of conduction band in the spin-down channel for both alloys has a major contribution from Co-*d* orbitals, whereas the main contribution of the spin-up channel is due to the strongly hybridized *d*-orbitals of Cr and Co atoms. It is also evident that the contribution of *s* and *p* orbitals of the main group elements Z=Al/Si is negligible at the top of the valence band and bottom of the conduction band in both spin channels.

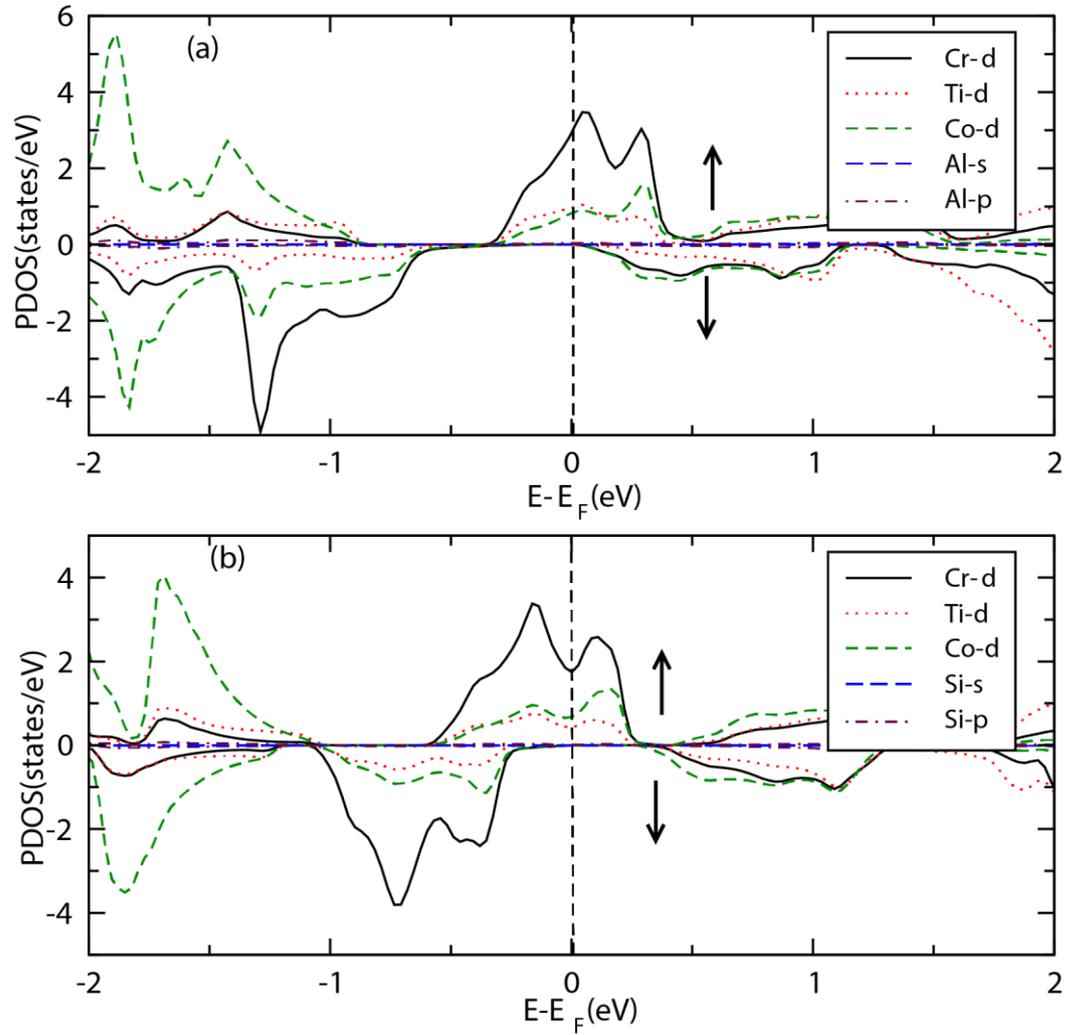

**Figure 5.** Projected density of states (PDOS) showing contribution from different atomic orbitals for (a) CrTiCoAl and (b) CrTiCoSi compounds.

The spin polarization can be calculated using the total density of states at the Fermi energy level ($E_f$) using the following relation:

$$P = \frac{\rho_{\uparrow}(E_f) - \rho_{\downarrow}(E_f)}{\rho_{\uparrow}(E_f) + \rho_{\downarrow}(E_f)} \times 100, \tag{14}$$

where, $\rho_{\uparrow}(E_f)$ and $\rho_{\downarrow}(E_f)$ stand for the total density of states at Fermi energy level of the majority and minority spin channels, respectively. As a result of the half metallic behavior of CrTiCoAl, it exhibits a 100% spin polarization (see Figure 2), whereas, CrTiCoSi exhibits a high spin polarization of 99.7%, see Table 5.

The total magnetic moment of both compounds as well as the local magnetic moment of every atom is listed in Table 5.

**Table 5.** Spin Polarization, magnetic moments and Curie temperature of CrTiCoAl and CrTiCoSi alloys.

|  | P(%) | $M_{total}$ ($\mu_B$) | $M_{Cr}$ ($\mu_B$) | $M_{Ti}$ ($\mu_B$) | $M_{Co}$ ($\mu_B$) | $M_Z$ ($\mu_B$) | $T_C$ (K) |
|---|---|---|---|---|---|---|---|
| CrTiCoAl | 100 | -2.00 | -1.83 | 0.17 | -0.24 | 0.01 | 385 |
| CrTiCoSi | 99.7 | -0.99 | -0.77 | 0.04 | -0.20 | 0.01 | 203 |

As shown in Table 5, the maximum contribution to the net magnetic moment comes from Cr atom (-1.83μB in CrTiCoAl alloy and -0.77μB in CrTiCoSi), whereas the main group elements, Si and Al with local magnetic moment of 0.01μB have a negligible contribution. Further, in both compounds, Ti is coupled antiferromagnetically to Cr and Co atoms, thus reducing the net magnetic moment, whereas Cr and Co atoms are coupled ferromagnetically. The total magnetic moment of CrTiCoAl obeys the linear relation as per Slater-Pauling rule[64, 65] given by equation 15.

$$M_{tot} = Z_{tot} - 24,$$ (15)

where, $M_{tot}$ is the next magnetic moment and $Z_{tot}$ is the total number of valence electrons. Additionally, the Curie temperature for both alloys is calculated as per relation 1.

$$T_c = 23 + 181 M_{total}$$ (16)

The Curie temperature has been computed in a similar manner for compounds CoYCrZ (Z = Si,Ge,Ga,Al)[30], CoFeCrZ (Z = P,As and Sb) [32], CoRhMnZ (Z = Al, Ga, Ge and Si)[33], VTiRhZ (Z = Si, Ge, and Sn)[62] and Co₂VAl[66]. As given in Table 4, Curie temperature for CrTiCoAl is above room temperature, whereas for CrTiCoSi it is 203K, which is below the RT.

## 4. Conclusion

Ab-initio calculations are performed to investigate the thermodynamic stability, dynamical phonon dispersion relations, mechanical properties, electronic band structure and magnetic properties of CrTiCoZ (Z= Al/Si) quaternary Heusler alloys. The thermodynamic calculations shows that Y-Type III is the most stable atomic configuration. This was further confirmed by studying the phonon dispersion relations, which shows only positive frequencies making them dynamically stable. The mechanical stability of these compounds was also checked along with calculating their melting temperatures, which are found well above the room temperature in both cases. The electronic structure calculations predict half-metallic behavior with a direct band gap in the spin-down channel for CrTiCoAl. However, CrTiCoSi alloy is found to be semi-metallic in the spin-down and metallic in the spin-up channels. The values of the net magnetic moment are -2μB and -0.99μB for CrTiCoAl and CrTiCoSi, respectively. The Curie temperature is well above room temperature for the case of CrTiCoAl, whereas it is below room temperature for CrTiCoSi. Thus, we find that only CrTiCoAl is suited for application as a spin injector in spintronics devices.